\documentclass[letter]{aa} % for the letters 
\usepackage{graphicx}
\usepackage{txfonts}
\usepackage{natbib}
\begin{document}
   \title{From high-mass starless cores to high-mass protostellar objects\thanks{{\it Herschel} is an ESA space observatory with science instruments provided by
Principal Investigator consortia. It is open for proposals for observing time
from the worldwide astronomical community.}}

%   \subtitle{I. Overviewing the $\kappa$-mechanism}

   \author{H.~Beuther
          \inst{1}
          \and
          Th.~Henning
           \inst{1}
          \and
          H.~Linz
          \inst{1}
          \and
          O.~Krause
          \inst{1}
          \and
          M.~Nielbock
          \inst{1}
          \and
          J.~Steinacker
          \inst{2,1}
           }
   \institute{$^1$ Max-Planck-Institute for Astronomy, K\"onigstuhl 17,
              69117 Heidelberg, Germany, \email{name@mpia.de}\\
             $^2$ LERMA \& UMR 8112 du CNRS, Observatoire de Paris, 61 Av. de l'Observatoire, 75014 Paris, France
       }

%   \date{Received September 15, 1996; accepted March 16, 1997}

% \abstract{}{}{}{}{} 
% 5 {} token are mandatory
%             \abstract{blablabla}

\abstract
{}
{Our aim is to understand the evolutionary sequence of high-mass star
  formation from the earliest evolutionary stage of high-mass starless
  cores, via high-mass cores with embedded low- to intermediate-mass
  objects, to finally high-mass protostellar objects.}
{{{\it Herschel}} far-infrared PACS and SPIRE observations are
  combined with existing data at longer and shorter wavelengths to
  characterize the spectral and physical evolution of massive
  star-forming regions.}
{The new {{\it Herschel}} images spectacularly show the evolution of the
  youngest and cold high-mass star-forming regions from mid-infrared
  shadows on the Wien-side of the spectral energy distribution (SED),
  via structures almost lost in the background emission around
  100\,$\mu$m, to strong emission sources at the Rayleigh-Jeans tail.
  Fits of the SEDs for four exemplary regions covering evolutionary stages
  from high-mass starless cores to high-mass protostellar objects
  reveal that the youngest regions can be fitted by single-component
  black-bodies with temperatures on the order of 17\,K. More evolved
  regions show mid-infrared excess emission from an additional warmer
  component, which however barely contributes to the total
  luminosities for the youngest regions. Exceptionally low values of
  the ratio between bolometric and submm luminosity additionally
  support the youth of the infrared-dark sources.}
{The {{\it Herschel}} observations reveal the spectral and physical properties
  of young high-mass star-forming regions in detail.  The data clearly
  outline the evolutionary sequence in the images and SEDs. Future
  work on larger samples as well as incorporating full radiative
  transfer calculations will characterize the physical nature at the
  onset of massive star formation in even more depth.}

%\titlerunning{}
   \maketitle

\section{Introduction}
\label{intro}

\begin{figure*}[ht]
\includegraphics[angle=-90,width=18.4cm]{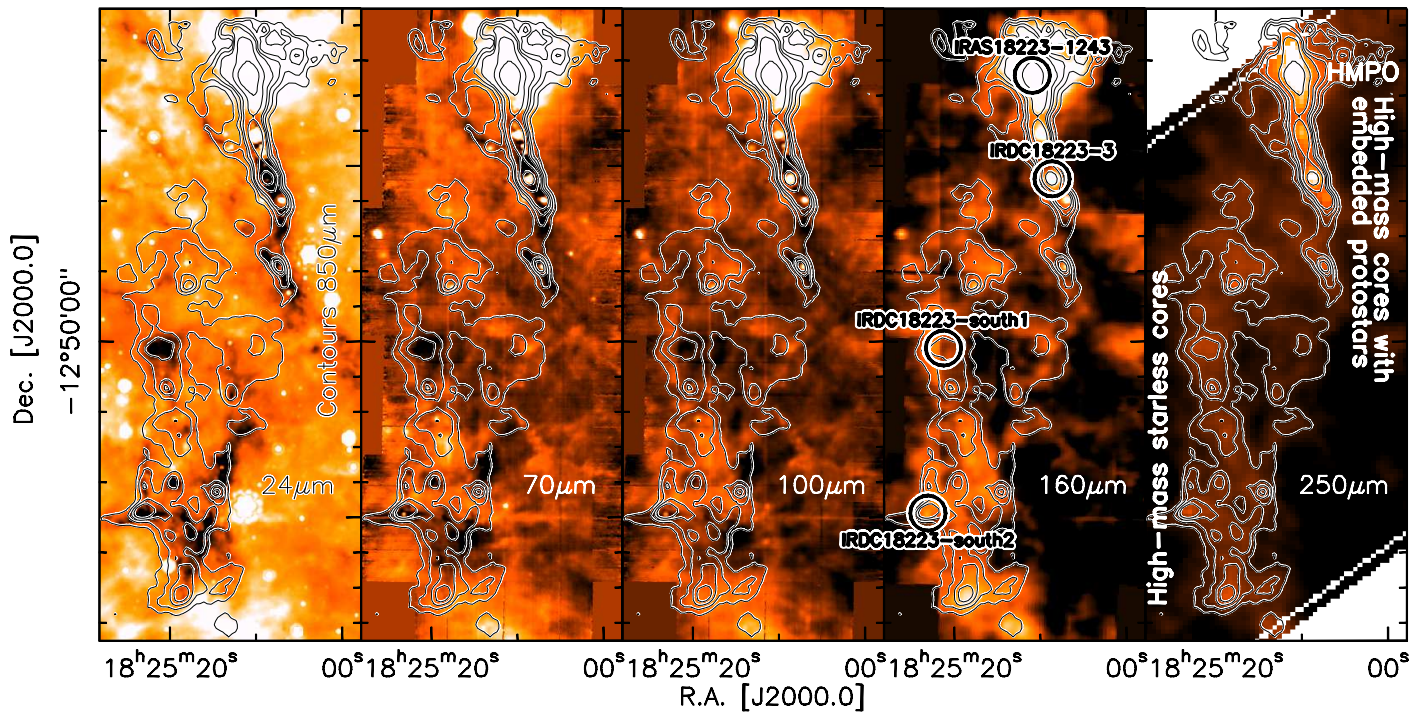}
\caption{The color-scale presents the wavelengths as marked in each
  panel. 70, 100 and 160\,$\mu$m are from PACS, 250\,$\mu$m is from
  SPIRE. The contours show the SCUBA 850\,$\mu$m continuum emission
  from 0.05 to 0.35\,Jy\,beam$^{-1}$ (in 0.1\,Jy\,beam$^{-1}$ steps)
  and additionally two stronger contours at 0.65 and
  1.25\,Jy\,beam$^{-1}$.  The right panel labels which part of the
  region are dominated by what evolutionary stage. The circles and
  labels in the 160\,$\mu$m panel mark the sources for which SEDs were
  extracted (Fig.~\ref{fig_seds}).}
\label{continuum}
\end{figure*}

Characterizing the early evolutionary stages of (high-mass) star
formation is difficult because the cold cores ($\sim$10 to 20\,K) have
the peak of the spectral energy distribution (SED) at far-infrared
wavelengths which were hardly accessible until recently.  {\it Herschel} has
changed this situation completely, and we are now able to study the
young (massive) star-forming regions in detail. For this early
{\it Herschel} study we selected the complex associated with the High-Mass
Protostellar Object (HMPO) IRAS\,18223-1243 and the Infrared Dark
Cloud IRDC\,18223-3 at a distance of $\sim$3.7\,kpc \citep{sridha}.
This region hosts most evolutionary stages in massive star formation,
starting with high-mass starless cores, continuing with
high-mass cores with embedded low- to intermediate-mass protostars
potentially forming massive stars, to HMPOs with already embedded and
likely still accreting massive protostars (Fig.~\ref{continuum}).

Previous studies focused either on the HMPO IRAS\,18223-1243 (e.g.,
\citealt{sridha,beuther2002a}) or at a high-mass core with embedded
low- to intermediate-mass protostar IRDC\,18223-3
\citep{garay2004,beuther2007a,fallscheer2009}.  However, the whole
complex with the youngest starless cores has not been studied so far.
Here we combine {\it Herschel} PACS and SPIRE data at far-infrared
wavelengths
\citep{A&ASpecialIssue-HERSCHEL,A&ASpecialIssue-PACS,A&ASpecialIssue-SPIRE}
with Spitzer mid-infrared data and (sub)mm continuum observations at
the Rayleigh-Jeans tail of the spectrum.

\section{Data and observations} 
\label{obs}

The cloud complex with a size of $\sim 4'\times 9'$ was observed
with PACS \citep{A&ASpecialIssue-PACS} on {\it Herschel}
\citep{A&ASpecialIssue-HERSCHEL} on 2009 October 9 within the
science demonstration program. Scan maps in two orthogonal directions
with scan leg lengths of $18'$ and $6'$, respectively, were
obtained with the medium scan speed of 20$''$/s. The raw data have
been reduced with the HIPE software, version 3.0, build 455
\citep{A&ASpecialIssue-PACS}.  Beside the standard steps leading to
level-1 calibrated data, a second-level deglitching as well as a
correction for offsets in the detector sub-matrices were
performed. Finally, the data were highpass-filtered, using a
median window of the size of the full scan legs, to remove
the effects of bolometer sensitivity drifts and the 1/f noise along
the course of the data acquisition.  We masked out emission structures
(visible in a first iteration) before computing and subtracting this
running median.  Thereby over-subtraction of source emission in the
highpass filtering step can be minimized. The flux correction factors
provided by the PACS ICC team were applied. The beam sizes of the
70, 100 and 160\,$\mu$m data are $\sim$5.6$''$, $\sim$6.8$''$ and
$\sim$11.4$''$, respectively. The 100\,$\mu$m PACS flux within an
aperture of $40''$ toward the IRAS source 18223-1243 is fully
consistent with the calibration uncertainties of both missions
($\sim$17\% measured difference).

Maps at 250, 350, and 500 $\mu$m were obtained with SPIRE
\citep{A&ASpecialIssue-SPIRE} on 2009 October 19. Two $14'$ scan legs
were used to cover the source. The data were processed within HIPE
with the standard photometer script up to level 1.  During baseline
removal, we masked out the high-emission area associated with the IRAS
source. Because no cross-scan data were obtained for these
observations, the iterative de-striping algorithm was invoked to
mitigate this effect \citep{A&ASpecialIssue-Bendo}.  The beam sizes at
250, 350 and 500\,$\mu$m are $\sim$18.1$''$, $\sim$24.9$''$ and
$\sim$36.6$''$, respectively. We estimate the fluxes to be accurate
within 20\%.

The MIPS 24 and 70\,$\mu$m data (from MIPSGAL, \citealt{carey2009}) as
well as the IRAC 3.5 to 8\,$\mu$m (from GLIMPSE,
\citealt{churchwell2009}) and the PdBI 3.2\,mm continuum data were
first presented in \citet{beuther2005d} and \citet{beuther2007a}. On
the long-wavelength side, we use the SCUBA 850\,$\mu$m data from the
SCUBA archive \citep{difrancesco2008} and the 1.2\,mm continuum flux
measurement observed with the IRAM 30\,m telescope
\citep{beuther2002a}. The accuracy of the flux measurements at (sub)mm
wavelength is estimated to be correct within $\sim$15\% and for the
24\,$\mu$m within 20\%. Because no sources are detected in the IRAC
bands for most of our targets (except of the HMPO), the 3$\sigma$
upper limits of the four Spitzer IRAC datasets are 0.05\,mJy at 3.6
and 4.5\,$\mu$m, 0.13\,mJy at 5.8\,$\mu$m and 0.15 at 8\,$\mu$m.

\section{Results}

\subsection{General structure of the complex}

Figure \ref{continuum} gives an overview of the whole complex from
mid-infrared via far-infrared to (sub)mm wavelengths. The covered
region of approximately $4'\times 9'$ corresponds at the given
distance to an area of $\sim$42\,pc$^2$ with a length of $\sim$10\,pc.
At 70\,$\mu$m, the linear spatial resolution is $\sim$20\,000\,AU.
Globally, one can clearly distinguish the different evolutionary
stages and the corresponding spectral behavior in these images.  We
will discuss the global properties first, spectral energy
distributions of exemplary sources will be presented in
Sect.~\ref{seds}.

While the HMPO and IRAS source in the north of the field
(IRAS\,18223-1243) is a strong emission source at all covered
wavelengths, this is clearly not the case for most other parts of the
region. For example, the filamentary structure south of
IRAS\,18223-1243 is a clear (sub)mm continuum emission region, but it
shows pronounced filamentary absorption structures against the bright
Galactic background at 24\,$\mu$m and shorter wavelengths (see also
\citealt{beuther2005d,beuther2007a}).  Nevertheless, while the (sub)mm
emission peaks in the filament extending south to a declination of
$\sim -12^o50'00''$ show only absorption in the IRAC bands up to
8\,$\mu$m wavelength \citep{beuther2005d}, they are weak emission
sources in the Spitzer 24\,$\mu$m band.  Combining this weak
mid-infrared emission features with outflow signatures like the
``extremely green objects'' (EGOs, \citealt{cyganowski2008}) and CO
and CS line wing emission favors the interpretation of these sources
as low- to intermediate-mass protostars embedded in high-mass cores
that will likely form massive stars at the end of the evolution
\citep{beuther2007a}. Again different from these sources are the (sub)mm
emission peaks found south of $\sim -12^o50'00''$. Almost all of these
(sub)mm sources are either extinction features at 24\,$\mu$m
wavelength or they show extended emission associated likely with the
fore- or background. None of these sources show clear 24\,$\mu$m
emission indicative of a protostar during its formation. While
some of these (sub)mm sources start appearing as far-infrared emission
peaks at 70\,$\mu$m in the {\it Herschel} data, others remain in absorption
or at least ``inconspicuous'' to wavelengths as long as 100\,$\mu$m.
While more targeted studies in the past almost always found signs of
star formation activity toward IRDCs (e.g.,
\citealt{rathborne2006,beuther2007g,motte2007}) less biased {\it Herschel}
studies now have the potential to identify and characterize genuine
high-mass starless cores.

From a more general point of view, these images can be considered as
templates of how young star-forming regions change their appearance
with wavelength: While the youngest and coldest regions appear as
shadows from near- to mid-infrared wavelengths against the stronger
diffuse background emission, background and source emission
approximately equalize at far-infrared wavelength (e.g., 100\,$\mu$m
panel in Fig.~\ref{continuum}). Going to even longer wavelengths, the
cold SEDs raise the emission of the cores considerably above the
background, and one clearly identifies them as emission sources.  This
evolutionary picture strongly resembles that of low-mass star-forming
regions (e.g., \citealt{andre1993}).

\subsection{Spectral energy distributions}
\label{seds}

\begin{figure*}[ht]
\includegraphics[angle=-90,width=5.0cm]{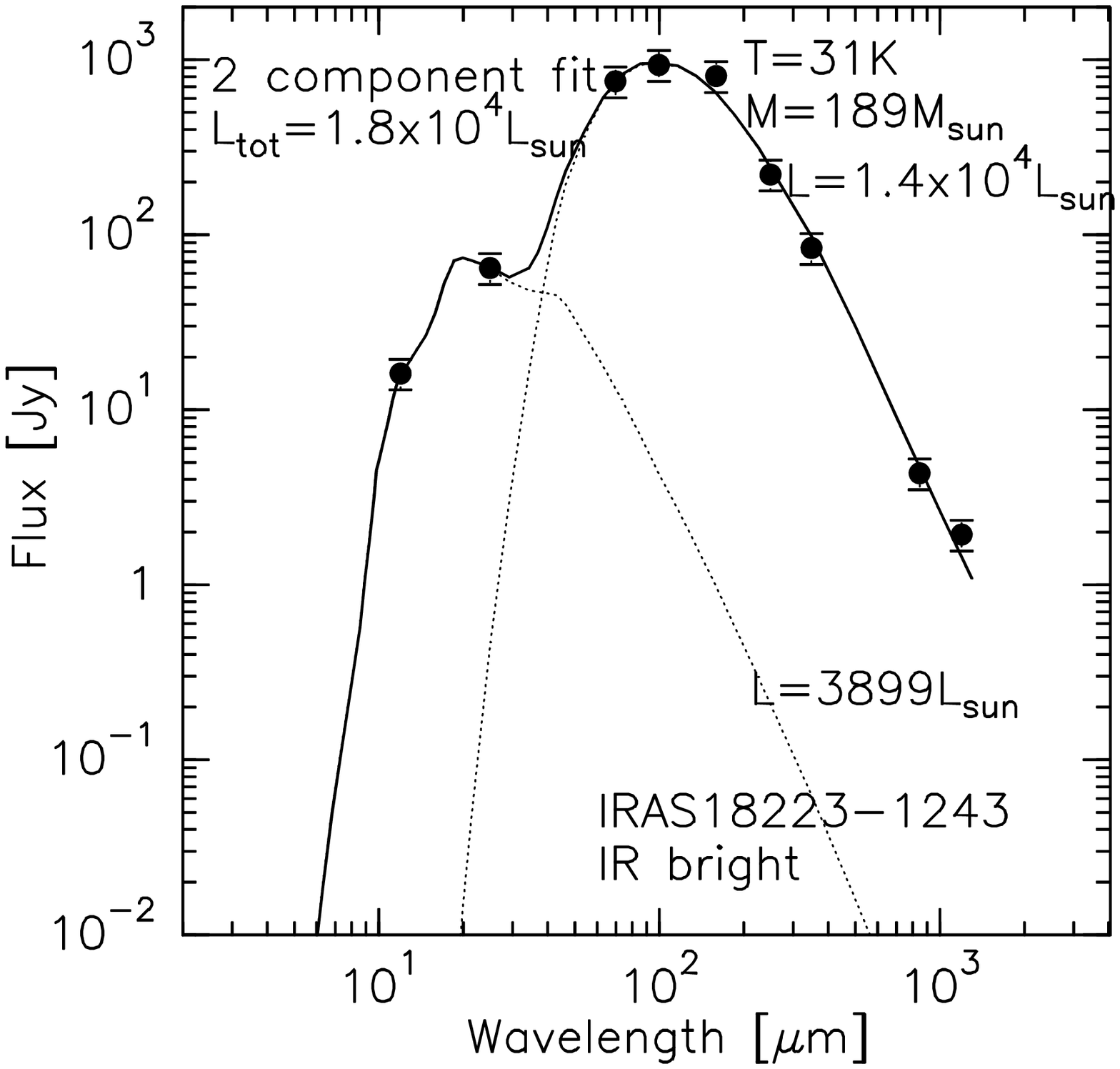}
\includegraphics[angle=-90,width=4.8cm]{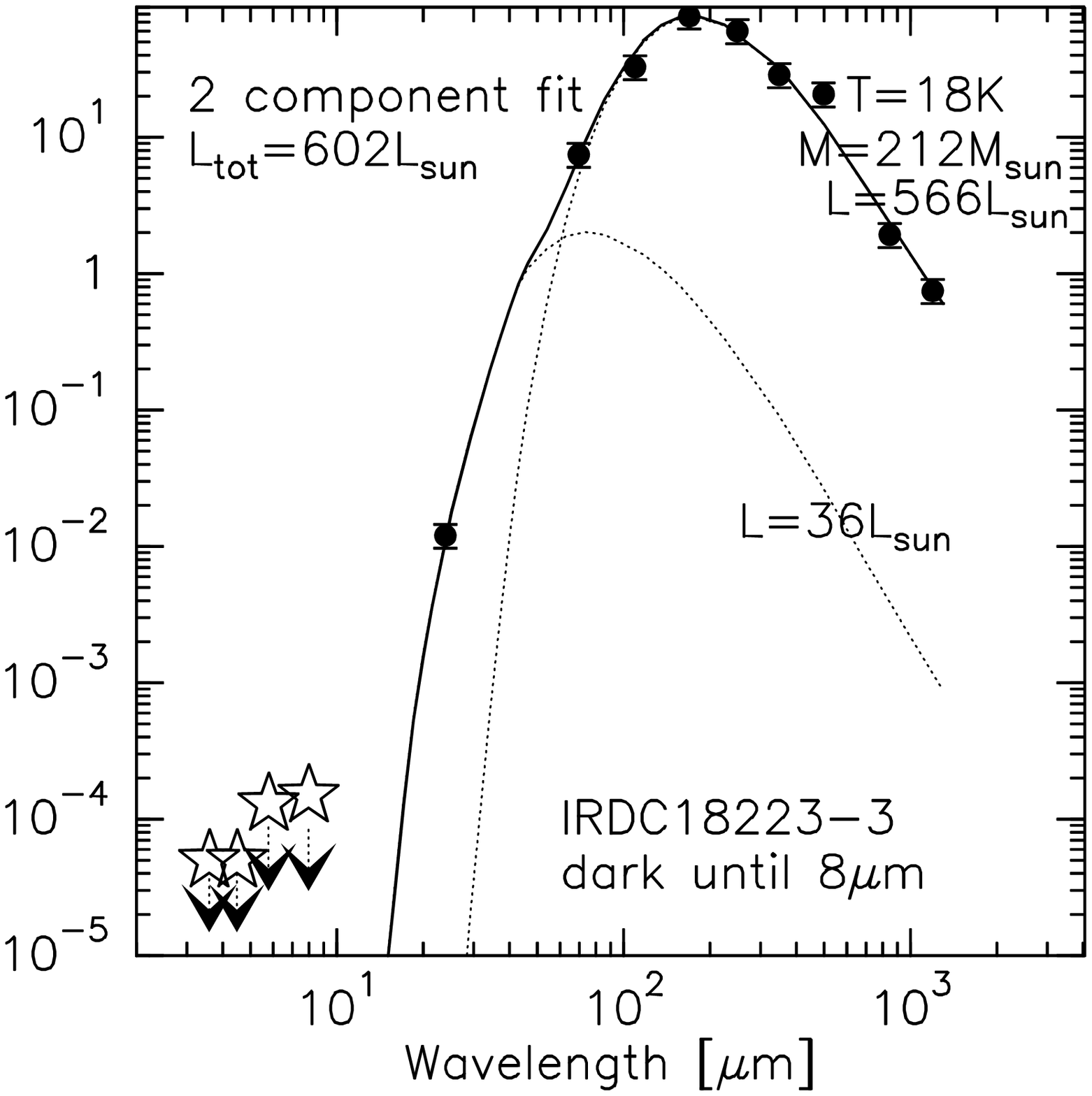}
\includegraphics[angle=-90,width=4.2cm]{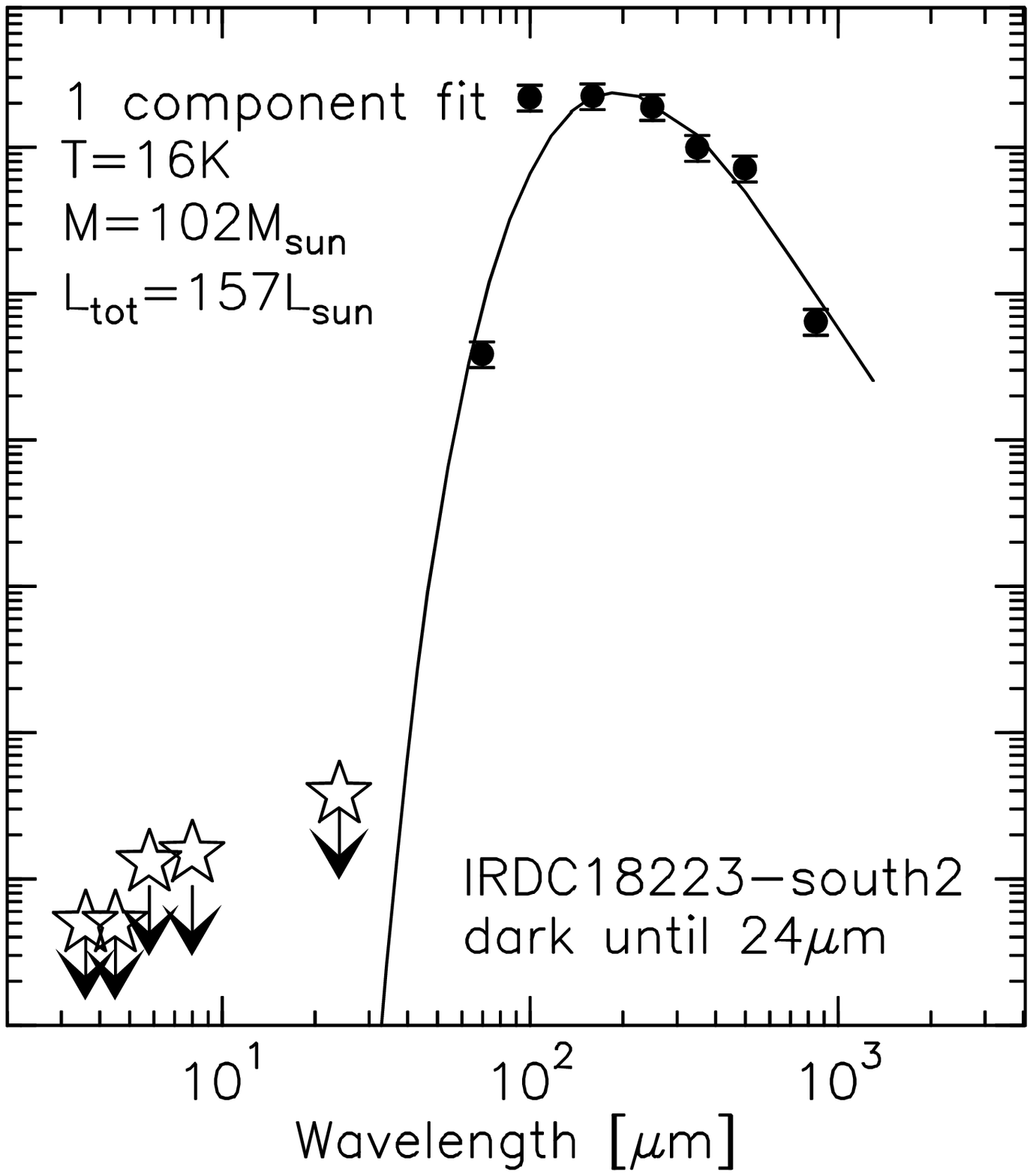}
\includegraphics[angle=-90,width=4.2cm]{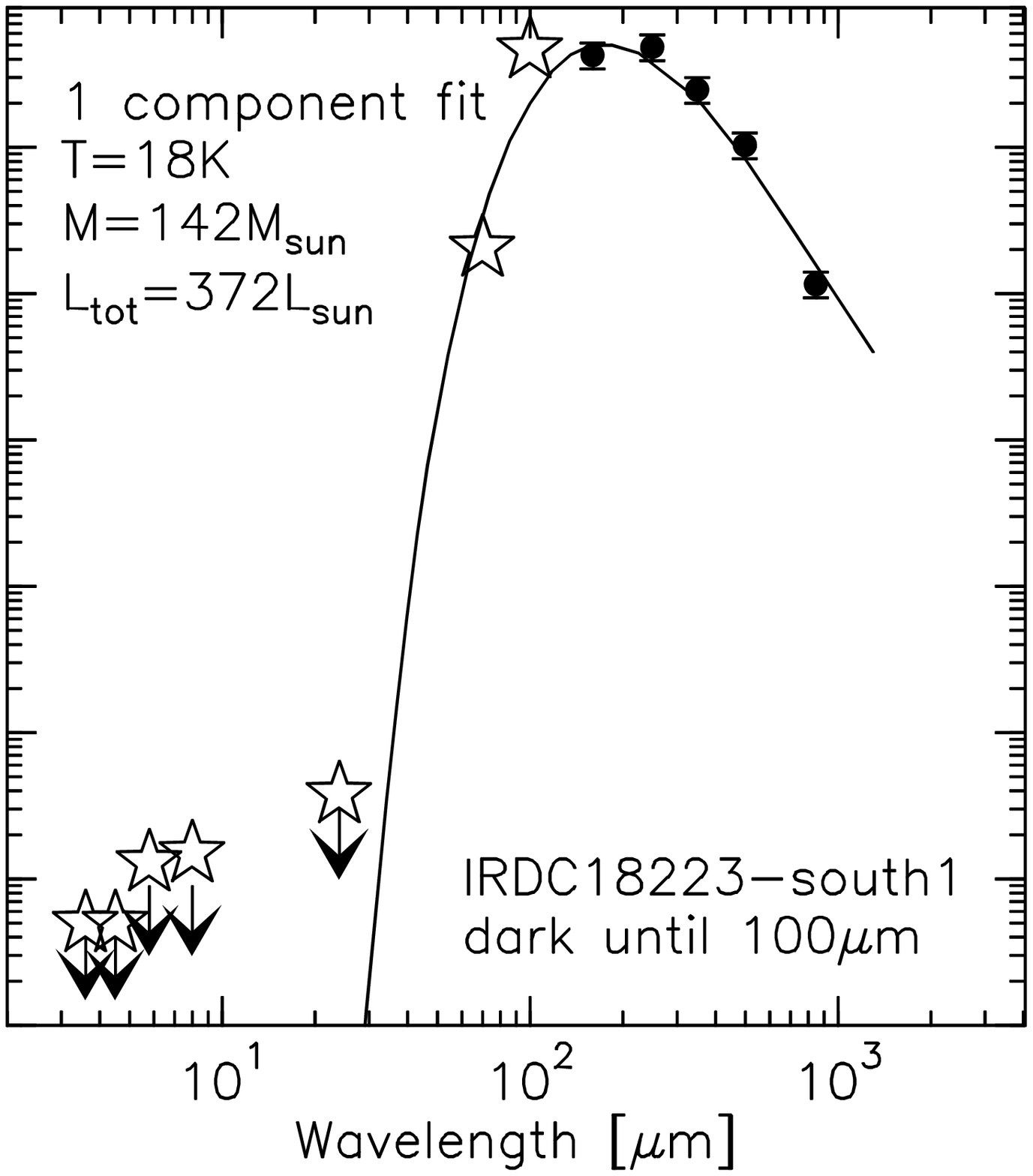}
\caption{SEDs of the four discussed example regions shown in
  Fig.~\ref{continuum}. The left panel presents an already more
  evolved HMPO, and going to the right, the regions become
  progressively younger with IRDC\,18223-south1 being only detected
  longward of 100\,$\mu$m. The stars with arrows mark upper limits,
  whereas stars only give estimates of the existing background flux.}
\label{fig_seds}
\end{figure*}

To better characterize the evolutionary sequence, we selected four
sources representing different evolutionary stages: (a) the HMPO
IRAS\,18223-1243 \citep{sridha,beuther2002a}, (b) the low- to
intermediate-mass protostar embedded in the infrared dark filament
IRDC\,18223-3 \citep{beuther2005d,beuther2007a}, (c) a 24\,$\mu$m
shadow in the south that becomes detectable from 70\,$\mu$m onward
(IRDC\,18223-south2), and (d) a source that remains dark up to
100\,$\mu$m (IRDC\,18223-south1). The sources are marked in
Fig.~\ref{continuum}, the peak positions in R.A.~and Dec.~(J2000.0)
are (a) 18:25:10.67 -12:42:25.8, (b) 18:25:08.61 -12:45:19.7, (c)
18:25:22.89 -12:54:50.7, (d) 18:25:19.15 -12:50:11.2, respectively.

%\begin{table}[ht]
%\caption{Source positions}
%\begin{tabular}{lrr}
%\hline \hline
% & R.A. & Dec. \\
% & [J2000.0] & [J2000.0] \\
%\hline
%18223-1243   & 18:25:10.67 & -12:42:25.8 \\
%18223-3      & 18:25:08.61 & -12:45:19.7 \\
%18223-south2 & 18:25:22.89 & -12:54:50.7 \\
%18223-south1 & 18:25:19.15 & -12:50:11.2 \\
%\hline \hline
%\end{tabular}
%\label{positions}
%\end{table}

Because {\it Herschel} fully covers the peak of the SED even for the coldest
regions, and is unique with respect to spatial resolution, sensitivity
and robustness against saturation at the given wavelengths, this
offers the opportunity to characterize this evolutionary sequence from
an SED point of view in detail.  Therefore we extracted the fluxes
over the full wavelength coverage between 12\,$\mu$m and 1.3\,mm
wavelength covering the Wien and Rayleigh-Jeans parts of the spectrum.
We derived the fluxes in all bands for uniform apertures.  For
IRDC\,18223-3 and IRDC\,18223-south2, the chosen aperture was based on
the coarsest spatial resolution of the SPIRE 500\,$\mu$m band
(FWHM$\sim$36.6$''$).  For the HMPO IRAS\,18223-1243, we chose a
$40''$ aperture to be able to combine the data with the IRAS fluxes at
12 and 25\,$\mu$m (Spitzer is saturated for that source at 24\,$\mu$m
and shorter wavelengths).  And for the youngest source we choose an
even larger aperture of $60''$ to cover the whole clump as visible in
Fig.~\ref{continuum}.  Table \ref{data} shows the derived fluxes and
Fig.~\ref{fig_seds} presents the corresponding SEDs.

\begin{table}[ht]
\caption{Source fluxes in [Jy]}
\begin{tabular}{lrrrr}
\hline \hline
& 18223-1243 & 18223-3 & 18223-south2 & 18223-south1 \\
\hline
$S_{12}$  & 16.2$^1$ & --       & --       & --   \\
$S_{24}$  & 64.9$^1$ & 12.1$^3$ & --       & --   \\
$S_{70}$  & 756.6    & 7.5$^4$  & 0.39$^5$ & --   \\
$S_{100}$ & 938.6    & 33.0     & 22.1$^6$ & --   \\
$S_{160}$ & 811.0    & 77.7     & 22.6$^6$ & 43.0 \\
$S_{250}$ & 222.0    & 60.6     & 19.1     & 48.9 \\
$S_{350}$ & 84.5     & 28.9     & 10.0     & 25.0 \\
$S_{500}$ & --$^2$   & 20.8     & 7.3      & 10.4 \\
$S_{850}$ & 4.4      & 1.9      & 0.65     & 1.2  \\
$S_{1200}$& 1.9      & 0.75     & --       & --   \\
\hline \hline
\end{tabular}
\label{data}
\footnotesize{The SCUBA 850\,$\mu$m data and the 1.2\,mm data are taken from \citet{difrancesco2008} and \citet{beuther2002a}, respectively. $^1$IRAS flux; $^2$Artifact (see Fig.~\ref{continuum}); $^3$\citet{beuther2007a}, $^4$Because the source is embedded in absorption filament, we selected smaller aperture just covering the emission source. $^5$Because of absorption close to the source, here we applied PSF photometry based on the template by \citet{A&ASpecialIssue-PACS}. $^6$We corrected for the mapping artifact visible around the source in Fig.~\ref{continuum}.}
\end{table}

The SEDs were fitted with modified Planck black-body functions
accounting for the wavelength-dependent emissivity of the dust. The
assumed dust composition follows \citet{ossenkopf1994}, and the
assumed gas-to-dust mass ratio is 100. The data longward of 70\,$\mu$m
can be well fitted with a single black-body function of cold dust and
gas (Fig.~\ref{fig_seds}). We also fitted the data for smaller
apertures ignoring the less resolved 350 and 500\,$\mu$m SPIRE data
points, and the derived temperatures agree well with our presented
fits.  Therefore, on the scales resolvable by these data for the cold
dust and gas components, single component fits are still adequate.
Furthermore, detections at 24\,$\mu$m and shorter wavelengths for
IRDC\,18223-3 and IRAS\,18223-1243 indicate additional inner heating
sources. This implies a temperature gradient throughout the inner core
for these more evolved regions. However, a sophisticated radiative
transfer modeling is out of the scope of this letter.  Although
nominally one can fit a second component to this warmer gas
(Fig.~\ref{fig_seds}), only the luminosities of these components are
useful parameters, while masses and temperatures are poorly
constrained because of rising optical depth at shorter wavelengths.
It should be noted that in particular for IRDC\,18223-3 this
additional luminosity only barely contributes to the total luminosity,
consistent with pre-{\it Herschel} fits \citep{beuther2007a}.  The gas
masses calculated from the fits using the \citet{ossenkopf1994} dust
model with thin ice mantles can be considered as lower limits.  Hence,
we also calculated the gas masses from the 850\,$\mu$m data assuming
optically thin dust continuum emission, a dust spectral index $\beta
=2$ (corresponding to a dust absorption coefficient $\kappa_{850\mu m}
\sim 0.8$\,cm$^2$g$^{-1}$) resembling the general ISM
\citep{hildebrand1983,hunter2000}, and the temperatures from the cold
component fits. Another quantity to differentiate evolutionary stages
is the ratio of bolometric to submillimeter luminosity $L/L_{\rm{sm}}$
that is usually applied to low-mass protostars but is also discussed in
the high-mass regime (e.g., \citealt{andre1993,mueller2002}). Table
\ref{results} lists the derived parameters.

\begin{table}[ht]
\caption{Parameters for the cold (c) and warm (w) components}
\begin{tabular}{lrrrrrrr}
\hline \hline
& $T_{\rm{c}}$ & $M_{\rm{c}}$ & $L_{\rm{c}}$ & $L_{\rm{w}}$ & $L_{\rm{sm}}$ & $\frac{L}{L_{\rm{sm}}}$ &  $M_{850}$ \\
& [K] & [M$_{\odot}$] & [L$_{\odot}$] & [L$_{\odot}$] & [L$_{\odot}$] & & [M$_{\odot}$] \\
\hline 

18223-12 & 31 & 189 & 13849  & 3898 & 155 & 115 & 452 \\
18223-3  & 18 & 212  & 566    & 36   & 54 & 11  & 413 \\
18223-s2 & 16 & 102  & 157    & --   & 21 & 8   & 169 \\
18223-s1 & 18 & 142  & 372    & --   & 36 & 10  & 249 \\
\hline \hline
\end{tabular}
\footnotesize{$T_{\rm{c}}$, $M_{\rm{c}}$, $L_{\rm{c}}$ and $L_{\rm{w}}$ are fit results. $L_{\rm{sm}}$ is the submm luminosity (longward of 400\,$\mu$m, and $M_{850}$ is the mass in the derived from only the 850\,$\mu$m flux (see main text).}
\label{results}
\end{table}

\section{Discussion and conclusion}
\label{general}

These regions represent a potential evolutionary sequence where
IRDC\,18223-south1 resembles the youngest detectable stage with a
large emission peak of cold dust continuum emission on the
Rayleigh-Jeans tail of the SED longward of 160\,$\mu$m. However, at
100\,$\mu$m the source is lost in the background, and at shorter
wavelengths it remains an absorption shadow. The source
IRDC\,18223-south2 appears a bit more evolved because one detects an
emission source from 70\,$\mu$m to longer wavelengths. Nevertheless,
the region is still young and cold without any emission at 24\,$\mu$m.
Continuing on the evolutionary ladder, IRDC\,18223-3 exhibits
24\,$\mu$m emission but remains dark in the IRAC bands.
\citet{beuther2007a} called this a high-mass core with an embedded
low- to intermediate-mass protostar potentially becoming massive at
the end of the evolution.  Finally, at the northern end of the
filament IRAS\,18223-1243 comprises all features of a typical HMPO
\citep{sridha}.

This sequence is also reflected in the SED fitting. For the cold
sources, the single black-body fits result in low temperatures between
16 and 18\,K, luminosities of only a few hundred L$_{\odot}$ and gas
masses of several hundred M$_{\odot}$.  The luminosities of these cold
sources are not related to internal heating, but are mainly produced
by the external radiation field.  The low temperatures are consistent
with previous estimates toward infrared dark clouds based on spectral
line observations (e.g., \citealt{sridharan2005,pillai2006}). We note
that our fitted temperatures are also in the range of temperatures
derived from COBE for the general ISM (e.g., \citealt{reach1995}). On
the given scales ($>$0.65\,pc) at these early evolutionary stages, the
gas and dust clumps still have temperatures comparable to those of the
general ISM. For the most evolved HMPO, the fitted cold temperature is
warmer than for the other sources.  Furthermore, the luminosity
already indicates the presence of an HMPO, while the gas reservoir is
still large, allowing for further accretion.  While the total
luminosity compares well with older fits that were only based on IRAS
data \citep{sridha}, the cold temperature of the new fit is lower than
those only based on IRAS data (28 versus 50\,K, respectively). This is
because IRAS alone did not sample the peak of the SED well and hence
overestimated the temperature. Therefore sampling the full SED is
important even for such evolved sources to derive reasonable
temperature estimates.

In the framework of the $L/L_{\rm{sm}}$ ratio, the HMPO is above 100
where 200 was defined as the border between class 0 and class I for
low-mass sources \citep{andre1993}. While this cannot easily be
translated to high-mass star-forming regions, it shows that even the
HMPO is still in a young evolutionary stage (see \citealt{sridha}).
The three infrared-dark sources all have exceptionally low $L/L_{\rm{sm}}$
ratios around 10, lower than the high-mass cores reported by
\citet{mueller2002}, and among the lowest values so far reported in
the literature (e.g., \citealt{young2003}). This supports the extreme
youth of these sources. That we do not see a marked
difference in $L/L_{\rm{sm}}$ for the three infrared-dark sources
indicates that -- in the evolution of the massive cores -- first
signatures for star formation do not come from a global warm-up of the
bulk of dust and gas. More evolved sources like the HMPO 18223-1243
in turn have multiple T components which show up at mid-infrared
wavelengths, but most of the cold dust can still be reasonably well
fitted by a higher but single temperature.

While we present here one of the first sources observed with PACS and
SPIRE, there is tremendous potential in this kind of observations. For
the future we anticipate full radiative transfer modeling of such
regions, which will result e.g.~in more detailed temperature maps
and density structures over the entire field. Furthermore, the
{\it Herschel} key project EPOS (PI O.~Krause) contains more than 40 regions
in young evolutionary stages. A combined analysis of the whole sample
should result in a robust characterization of the physical properties
and early evolution of high-mass star formation.

\begin{acknowledgements} 
  PACS has been developed by a consortium of institutes led by MPE
  (Germany) and including UVIE (Austria); KUL, CSL, IMEC (Belgium);
  CEA, OAMP (France); MPIA (Germany); INAF-IFSI/ OAA/OAP/OAT, LENS,
  SISSA (Italy); IAC (Spain).  This development has been supported by
  the funding agencies BMVIT (Austria), ESA-PRODEX (Belgium), CEA/CNES
  (France), DLR (Germany), ASI (Italy), and CICT/MCT (Spain). 
%SPIRE
%  has been developed by a consortium of institutes led by Cardiff
%  University (UK) and including Univ.  Lethbridge (Canada); NAOC
%  (China); CEA, LAM (France); IFSI, Univ. Padua (Italy); IAC (Spain);
%  Stockholm Observatory (Sweden); Imperial College London, RAL,
%  UCL-MSSL, UKATC, Univ. Sussex (UK); and Caltech, JPL, NHSC, Univ.
%  Colorado (USA). This development has been supported by national
%  funding agencies: CSA (Canada); NAOC (China); CEA, CNES, CNRS
%  (France); ASI (Italy); MCINN (Spain); Stockholm Observatory
%  (Sweden); STFC (UK); and NASA (USA).
\end{acknowledgements}

%\bibliography{/home/beuther/tex/bibliography}   
%\bibliography{/Users/henrikbeuther/paper/bibliography}
%\bibliographystyle{aa}    % this does the style, aa.bst necessary

\end{document}